\documentclass[preprint,showkeys,secnumarabic,amsfonts,showpacs,amsmath,amssymb]{revtex4}
\usepackage[dvips]{color}
\usepackage{epsfig}
\usepackage{amsmath}
\usepackage{graphicx}

\begin{document}
\title{Cosmic Dynamics in the Chameleon Cosmology}

\author{H. Farajollahi}
\email{hosseinf@guilan.ac.ir} \affiliation{Department of Physics,
University of Guilan, Rasht, Iran}
\author{A. Salehi}
\email{a.salehi@guilan.ac.ir} \affiliation{Department of Physics,
University of Guilan, Rasht, Iran}

\date{\today}

\begin{abstract}
 \noindent \hspace{0.35cm}

We study in this paper chameleon cosmology applied to Friedmann-Robertson-walker space, which gives
rise to the equation of state (EoS) parameter larger than $-1$ in the past and less than $-1$ today, satisfying current
observations. We also study cosmological constraints on the model using the
time evolution of the cosmological redshift of distant sources which directly probes the expansion history of the universe.
Due to the evolution of the universe's expansion rate, the model independent Cosmological Redshift Drift (CRD)test is expected to experience a small, systematic
drift as a function of time. The model is supported by the observational data obtained from the test.

\end{abstract}


\maketitle

\section{introduction}

The recent astrophysical data indicate that there is a dark energy (DE) providing approximately two third of the current universe
energy density which explain the current cosmic acceleration \cite{Reiss}\cite{Bennet}. The most obvious candidate to explain DE is the cosmological constant which can fit observations well. However it is so small ( of order $10^{-33}eV$), in comparison with the Planck scale ($10^{19}GeV?$) that suffer from fine-tuning and the coincidence problems \cite{Mohapatra}. Numerous other DE models are produced by some exotic matter like phantom (field with negative energy) or some other
(usually scalar ) matter \cite{Caldwell}. Unfortunately, such scalar field is usually very light and its coupling to matter should be tuned to
extremely small values in order to be compatible with the Equivalence Principle. In a sense, the cosmological evolution of
 the scalar field contradicts with the solar system tests \cite{Nojiri}.

From particle physics point of view, there is also wide-spread interest in the possibility that, in addition
to the matter described by the standard model, our universe may be populated by one or more scalar fields. This
might be a feature in high energy physics beyond the standard model and are often related to the presence of extra-dimensions
\cite{Brax}. The existence of scalar field may also explain the early and late time acceleration of the
universe \cite{Lyth}--\cite{Easson}. It is most often the case that such fields interact with matter; directly due to a Lagrangian
coupling , indirectly through a coupling to the Ricci scalar or as the result of quantum loop corrections \cite{Damouri}--\cite{Biswass}. If the
scalar field self-interactions are negligible, then the experimental bounds on such a field are very strong; requiring it to either
couple to matter much more weakly than gravity does, or to be very heavy \cite{Uzan}--\cite{Damourm}. Recently, a novel scenario was presented
 that employed self-interactions of the scalar-field to avoid the most restrictive of the current bounds \cite{Khourym}. In
the proposed  model, a scalar field couples to matter with gravitational strength, in harmony with general expectations from
string theory whilst at the same time remaining very light on cosmological scales. In this paper we will go further and,
contrary to most expectations presented in \cite{Khourym}, allow scalar field which is very light on cosmological scales,
to couple to matter much more strongly than gravity does, and yet still satisfies the current experimental and observational
constraints. The cosmological value of such a field evolves over Hubble time-scales and may cause the late-time acceleration of our universe \cite{Brax2}. The crucial feature these models possess is the mass dependency of their  scalar field on the
local background matter density. On the earth where the density is about $10^{30}$  times higher than the cosmological background, the Compton
wavelength of the field is small enough to satisfy all existing tests of gravity \cite{Mota1}. On the other hand, in the solar system, where the density is
several orders of magnitude smaller, the Compton wavelength of the field can be much larger \cite{Dimopoulos}. This means that, in these models, the scalar field may have a mass in the solar system much smaller than was allowed. In the
cosmological scale, the field is lighter with its energy density evolves slowly over cosmological time-scales and it may be considered as an
effective cosmological constant. Although
 the idea of a density-dependent mass term is not new \cite{Wett}--\cite{Mot}, the
work presented in \cite{Khourym} \cite{Brax2} is outstanding in that the scalar field can couple directly to matter with gravitational strength.

There are cosmological models \cite{Feng} were initially proposed to obtain a model of dark energy with the EoS parameter $\omega>-1$ in the past and
$\omega <-1$ at present. These models can be viewed as dynamical model for dark energy with
the feature that their EoS parameters can smoothly cross over the cosmological constant barrier $\omega =-1$ \cite{Gannouji} \cite{Andrianov}. To construct the model
it is necessary to add extra degrees of freedom with un-conventional features to the conventional single field
theory if we expect to realize viable models in the framework of gravity theory \cite{Zhao}. We believe that the chameleon model investigated in this work is capable to do the job. It predicts $\omega$ crossing scenario in the history of the universe in addition to the bouncing behaviour of the universe.

Since the cosmic acceleration affects the expansion
history of the universe, to understand the true nature of
the driving force, mapping of the cosmic expansion of the universe
is very crucial \cite{Linder}. Hence, we require various observational probes
in different redshift ranges to understand the expansion history of
the universe. The observational tools for probing the cosmic
acceleration broadly fall into two categories. Geometrical probe that deals with large scale
distances and volume which include luminosity distance measurements
of SNe Ia, angular diameter distance from first CMB acoustic peak,
baryon acoustic oscillations and so on. Dynamical probe that investigates
the growth of matter density perturbations to give rise to the
large scale structure such as galaxies and clusters of galaxies in
the universe. Using supernova as standard candles is a popular
method of constraining the properties of dark energy. Although these
methods are very simple and useful in constraining the various dark
energy models, at present, the luminosity distance measurements
suffer from many systematical uncertainties such as extinction by
dust and gravitational lensing \cite{Nordin}. Measuring
the expansion history from growth of matter perturbations also has
its limitations. It requires prior information of exact value of
matter density, initial conditions, cosmological model and so on \cite{Nordin}\cite{Huterer}.
 So the question arises, in any probe which is simple, depends on
fewer priors and assumptions.

One possible probe is "Cosmological
Redshift Drift" (CRD) test which maps the expansion of the universe
directly. It assumes that the universe is
homogeneous and isotropic at the cosmological scales \cite{Lis}. The CRD test while is based on very simple and straightforward
physics, observationally it is a very challenging task and
requires technological breakthroughs \cite{Cristiani}. This probe measures the dynamics of the
universe directly via the Hubble expansion factor. The time drift of the cosmological redshift
probes the universe in the redshift up to $z = 5$, whereas the other
cosmological tests based on SNeIa, BAO, weak lensing or number counts
of clusters have not advanced beyond $z = 2$. Its other
advantage is that it has controlled systematical
uncertainties and evolutionary effects of the sources \cite{Jain}. The test was first
proposed by the author in \cite{Sandage} in terms of the scale factor. So, the universe expansion rate is directly measured by the time evolution of
the scale factor, or change in redshift, $\dot{z}$. The measurement of the redshift of an
object today, is different from its measured value
after a time interval of several years. As explained in \cite{McV} the redshift drift signal, $\dot{z}$, is very small. In \cite{Loeb}, author was the first to suggest the possibility of measuring the
redshift drift by observing Lyman $\alpha$ ($LY\alpha$) absorption lines in the
spectra of quasars (QSOs), which reinforcs the importance of
this probe.

This work is arranged as follows. In the next section we present the chameleon model. We also
investigate the conditions for the EoS parameter in the model to across $-1$ and moreover examine the possible bouncing of the universe.
In section three, by using the field equations we introduce the observational data and describe their inclusion in our analysis through driving CRD test for the model. We also revisit the Chevallier-Polarski-Linder (CPL) model in a comparison with our model. In Section four, we present the summary and remarks.

\section{The Model, Phantom crossing and bouncing universe}

We consider the chameleon gravity in the presence of cold dark matter with  the action given by,
\begin{eqnarray}\label{action}
S=\int[\frac{R}{16\pi
G}-\frac{1}{2}\phi_{,\mu}\phi^{,\mu}+V(\phi)+f(\phi){\cal L}_{m}]\sqrt{-g}dx^{4},
\end{eqnarray}
where $R$ is Ricci scalar, $G$ is the newtonian constant gravity
and $\phi$ is the chameleon scalar field with the potential
$V(\phi)$. Unlike the usual Einstein-Hilbert action, the matter
Lagrangian ${\cal L}_{m}$ is modified as $f(\phi){\cal L}_{m}$, where $f(\phi)$ is
an analytic function of $\phi$. The last term in the Lagrangian brings about the nonminimal
interaction between the cold dark matter and chameleon field.
The variation of action (\ref{action})  with respect to the metric tensor components in a spatially flat FRW  cosmology
yields the field equations,
\begin{eqnarray}\label{fried1}
3H^{2}=\rho_{m}f+\frac{1}{2}\dot{\phi}^{2}+V(\phi),
\end{eqnarray}
\begin{eqnarray}\label{fried2}
2\dot{H}+3H^2=-\gamma\rho_{m}f-\frac{1}{2}\dot{\phi}^{2}+V(\phi),
\end{eqnarray}
where we put  $8\pi G=c=\hbar=1$ and assume a perfect fluid with $p_{m}=\gamma\rho_{m}$  . The energy density $\rho_{m}$ stands for the contribution
from the cold dark matter to the energy density. Also
variation of the action (\ref{action}) with respect to the scalar field  $\phi$ provides the wave
equation for chameleon field as
\begin{eqnarray}\label{phiequation}
\ddot{\phi}+3H\dot{\phi}=-V^{'}-\gamma\rho_{m}f^{'},
\end{eqnarray}
where prime indicated differentiation with respect to $\phi$.
From equations (\ref{fried1}), (\ref{fried2}) and (\ref{phiequation}), one can easily arrive at the  relation (extended conservation equation)
\begin{eqnarray}
\dot{(\rho_{m}f)}+3H\rho_{m}(1+\gamma)f=\gamma \rho_{m} \dot{\phi}f^{'},
\end{eqnarray}
which readily integrates to yield
\begin{eqnarray}
\rho_{m}=\frac{M}{f^{(1-\gamma)}a^{3(1+\gamma)}},
\end{eqnarray}
where $M$  is a constant of integration.
From equations (\ref{fried1}) and (\ref{fried2}) and in comparison with the standard friedmann equations we identify $\rho_{eff}$ and $p_{eff}$ as
\begin{eqnarray}\label{roef}
\rho_{eff}\equiv\rho_{m}f+\frac{1}{2}\dot{\phi}^{2}+V(\phi),
\end{eqnarray}
\begin{eqnarray}\label{pef}
p_{eff}\equiv\gamma\rho_{m}f+\frac{1}{2}\dot{\phi}^{2}-V(\phi),
\end{eqnarray}
with an effective equation of state, $p_{eff}=\omega_{eff}\rho_{eff}$. From equations (\ref{roef}) and(\ref{pef}), we yield,
\begin{eqnarray}\label{phi2}
\dot{\phi}^{2}=\frac{2}{\omega_{eff}-1}[\rho_{m}f(\gamma-\omega_{eff})-V(\omega_{eff}+1)].
\end{eqnarray}
The possibility of phantom crossing can
be discussed in here. If crossing happens at
time $t=t_{cross}$, from equation (9)  we have
\begin{eqnarray}\label{phi2t}
\dot{\phi}^{2}(t_{cross})=-(\gamma+1)\rho_{m}f.
\end{eqnarray}
Equation (\ref{phi2t}) shows that, for phantom crossing,  $f$ must be negative. Also equations (\ref{roef}) , (\ref{pef}) and (\ref{phi2}) show that, if  $\dot{\phi}^{2}>-(1+\gamma)\rho_{m}f$  then  $\omega_{eff}<-1$  and  if  $\dot{\phi}^{2}<-(1+\gamma)\rho_{m}f$ then $\omega_{eff}>-1$.

Now in order to close the system of equations we make the following ansatz:
We take the potential to be power law in $\phi$ as $V(\phi)=V_{0}\phi^{n}$ and assume $f$ behaves exponentially as $ f(\phi)=f_{0} e^{b\phi}$ where $n, b, V_{0}$ and $f_{0}$ are arbitrary constants. Figure (1a) shows that  the possible crossing happens only when $f$ is always negative. In order to keep $f$ negative a runaway behavior is considered for it which also satisfies equation (\ref{phi2t}). In figure (1b), $\dot{\phi^{2}}$ never intersect $-(1+\gamma)\rho_{m}f$ as $f$ selected to be always positive so phantom crossing does not occur.\\

\begin{tabular*}{2. cm}{cc}
\includegraphics[scale=.35]{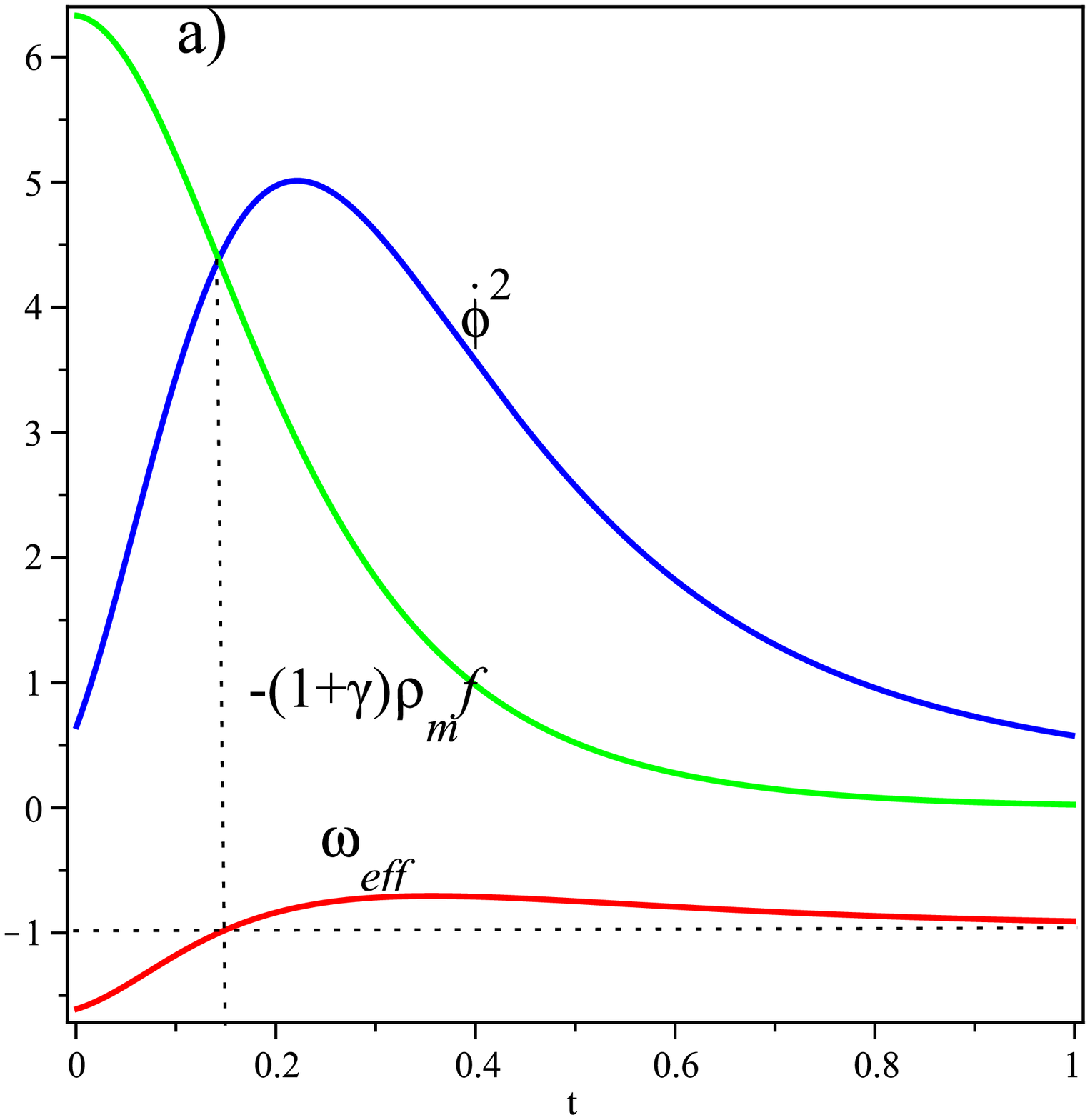}\hspace{.5 cm}
\includegraphics[scale=.35]{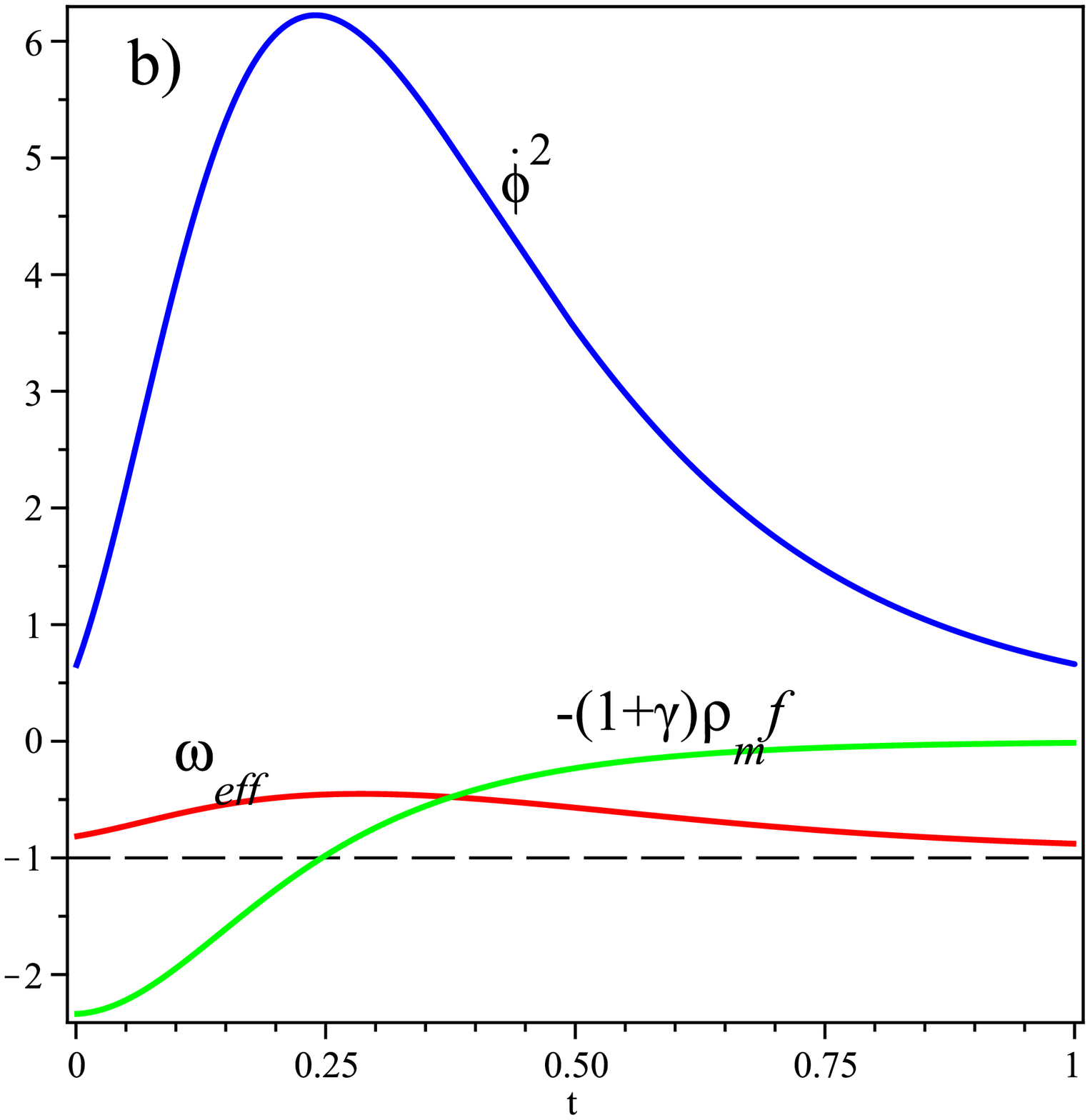}\\
Fig.1: Plots of $\omega_{eff}$, $\dot{\phi}^{2}$ and $-(1+\gamma)\rho_{m}f$ with $f={\it f_{0}}\,{{\rm e}^{b\phi}
 }$and $V=V_{0}\phi^{n}$,($b = -1$ ,$n=-1$)\\
for a) $f<0$ with $(f_{0}=-7) $ and  b) $f>0$ with $(f_{0}=2)$. \\Initial values are $\phi(0)=1$,
$\dot{\phi}(0)=-0.8$, $a(0)=0.8$ and $\dot{a}(0)=0.1$.
\end{tabular*}\\

Alternatively, we may take $f$ to be power law in $\phi$, $f(\phi)=f_{0}\phi^{n}$, and potential behaves exponentially, $ V(\phi)=V_{0} e^{b\phi}$. There are no priori physical motivation for these choices, so it is only purely phenomenological which leads to the desired behavior of phantom crossing.

In figure (2a), we examine phantom crossing for exponential $f$ and power law potential. We see that, as expected for $f<0$ we obtain crossing and for $f>0$ do not. At $f=0$ that $\dot{\phi}^{2}$ vanishes, from equation (\ref{phi2t}) and figure (2), $\omega_{eff}$ just becomes tangent to the line $-1$. The same argument can be made in figure (2b) for power law $f$ and exponential potential.\\

\begin{tabular*}{2. cm}{cc}
\includegraphics[scale=.35]{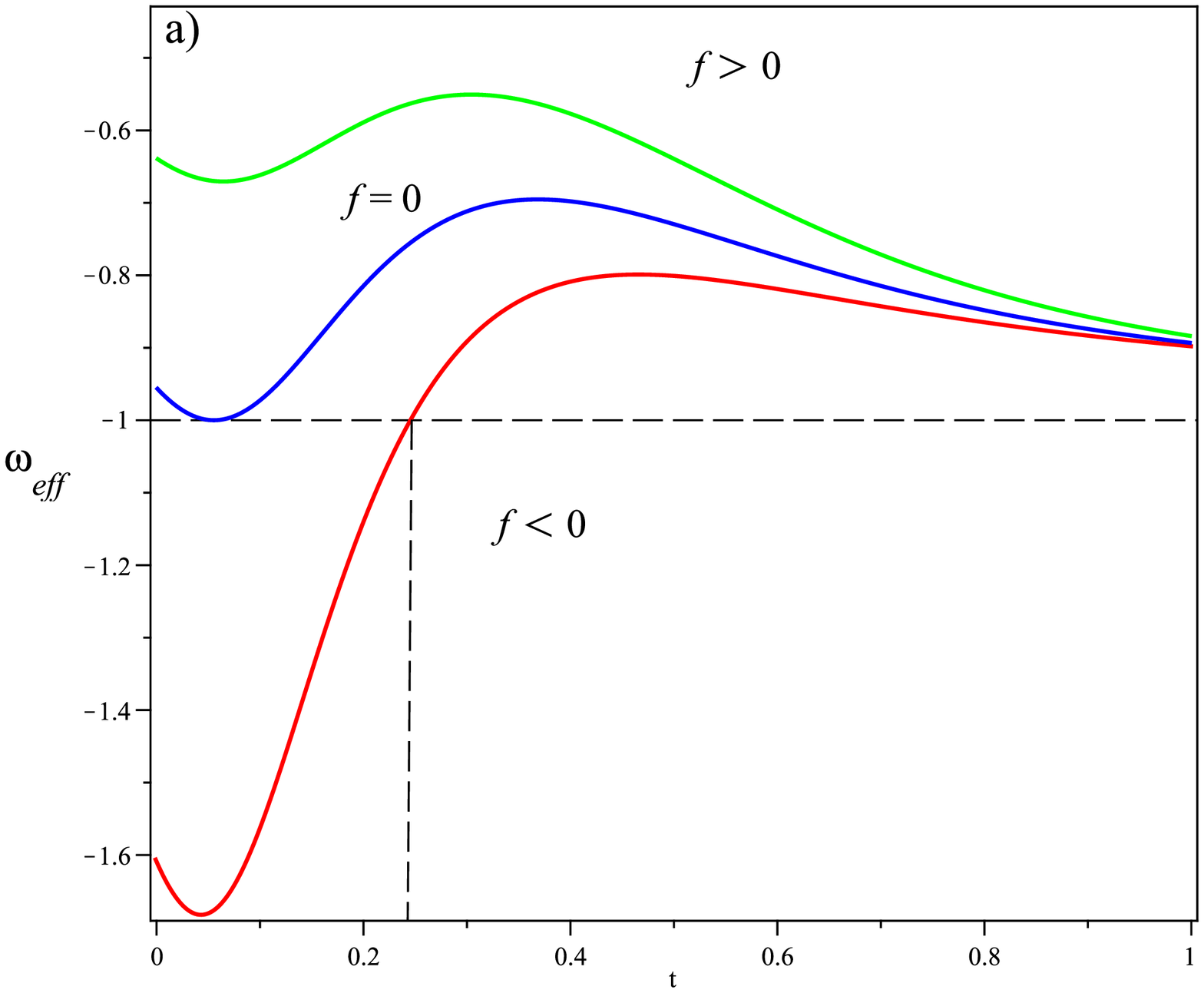}\hspace{.5 cm}
\includegraphics[scale=.35]{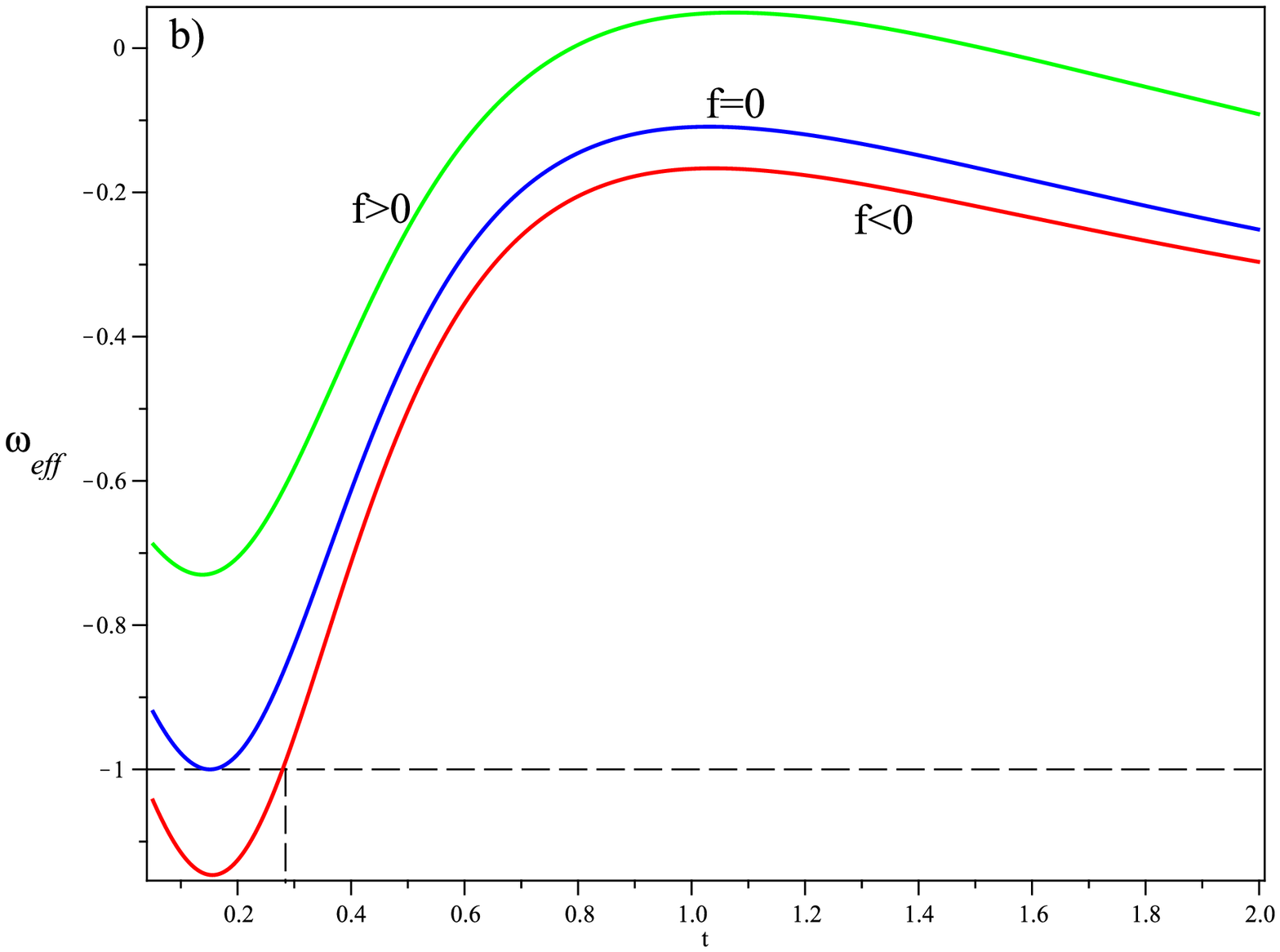}\\
Fig.2: Plot of $\omega_{eff}$ with a) $f={\it f_{0}}\,{{\rm e}^{b\phi}
 }$and $V=V_{0}\phi^{n}$ or b) $V={\it V_{0}}\,{{\rm e}^{b\phi
 }}$and $f=f_{0}\phi^{n}$
 \\for fixed $V_{0}$ and three different value of $f_{0}=-7$ ,$ f_{0}=0$ and $f_{0}=2$,($b = -1$ ,$n=-1$).\\ Initial values are $\phi(0)=1$,
$\dot{\phi}(0)=-0.8$, $a(0)=0.8$ and $\dot{a}(0)=0.1$.
\end{tabular*}\\

By choosing $t\approx 0$ to be the bouncing point, the solution for $a(t)$ and $H(t)$, shown in figure (3), provides a dynamical universe with
 contraction for $t<-0.02$, bouncing at $t=-0.02$ and then expansion for $t>-0.02$.

\begin{tabular*}{2. cm}{cc}
\includegraphics[scale=.35]{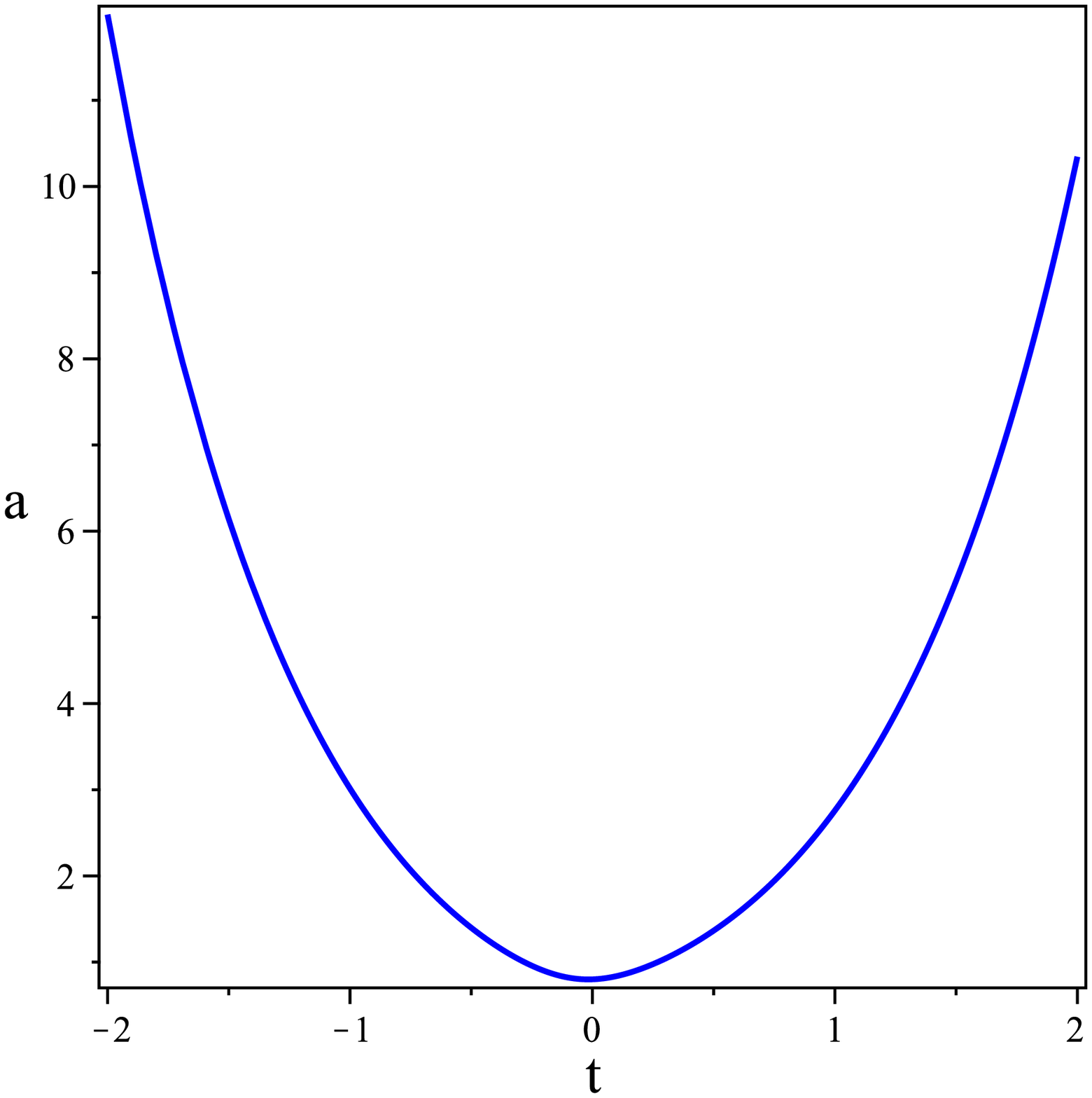}\hspace{.5 cm}
\includegraphics[scale=.35]{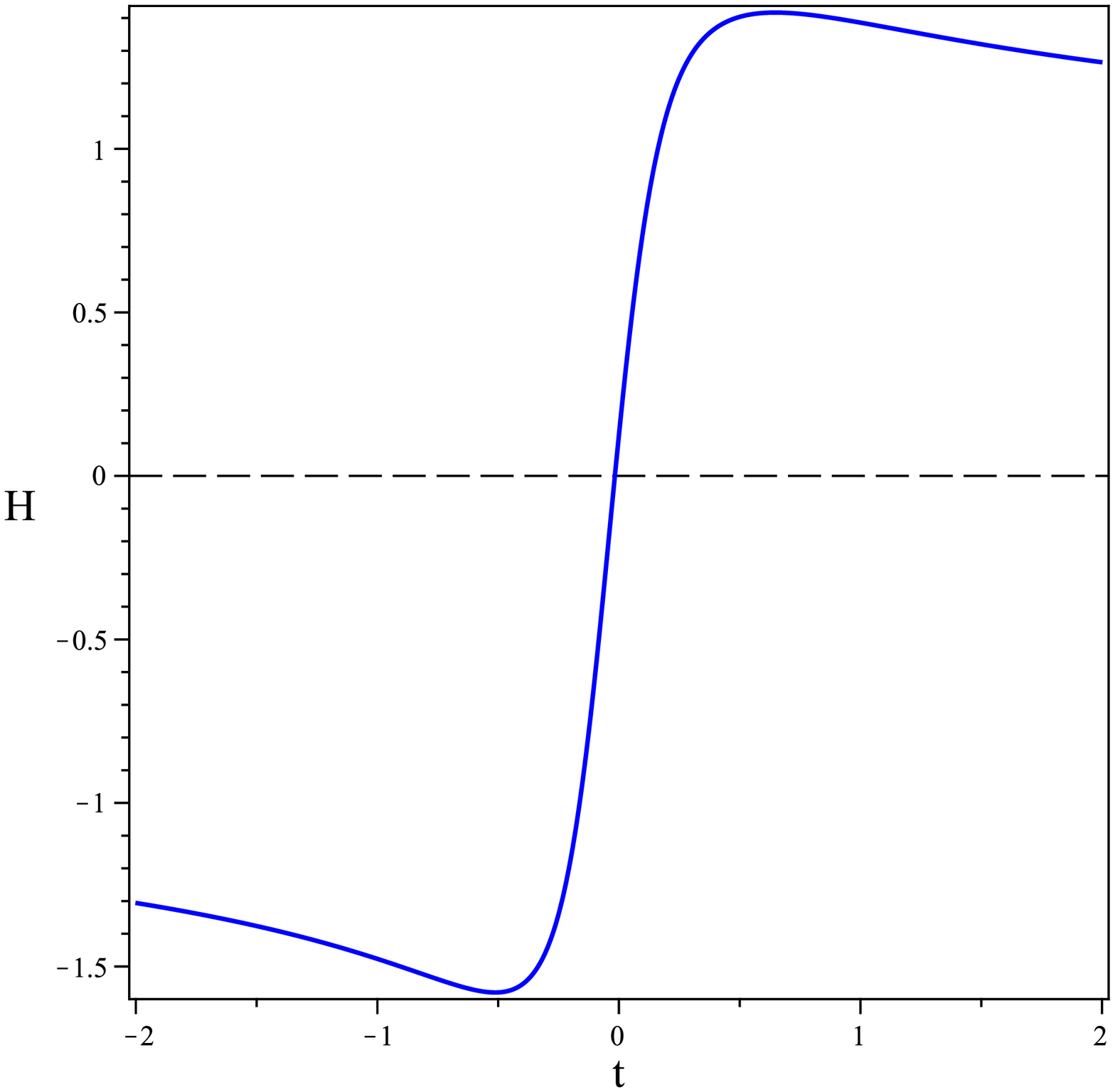}\\
Fig.3: Plots of   scalar factor $a(t)$ and
Hubble parameter $H$, at the bounce for\\ ${\it f_{0}}\,{{\rm
e}^{b\phi}}$, $f_{0} = -7$
 and $b =-1$. Initial values
are $\phi(0)=1$, $\dot{\phi}(0)=-0.8$, $a(0)=0.8$ and  $\dot{a}(0)=0.1$
.
\end{tabular*}\\

A detailed examination on the necessary conditions requires
for a successful bounce shows that during the contracting phase, the scale
factor $a(t)$ is decreasing, i.e., $\dot{a} < 0$, and in the
expanding phase we have $\dot{a} > 0$. At the bouncing point,
$\dot{a} = 0$, and so around this point $\ddot{a} > 0 $ for a
period of time. Equivalently in the bouncing cosmology the Hubble
parameter $H$ runs across zero from $H < 0$ to $H > 0$ and $H = 0$
at the bouncing point. A successful bounce requires that the following condition should be satisfied around bouncing
point,
\begin{eqnarray}\label{hdot1}
\dot{H}=-\frac{1}{2M^{2}_{p}}(1+\omega)\rho>0.
\end{eqnarray}
From figure (2a) and (2b) for $f<0$, we see that at $t\rightarrow 0$, $\omega_{eff}$ becomes less than $-1$ and $\dot{H}$ is positive which satisfies the above
condition. Also we see that at the bouncing point where the scale factor $a(t)$ is not zero we avoid singularity
faced in the standard cosmology.

\section{cosmological redshift-drift test(CRD)}
 The observed redshift of a distant source is given by
\begin{equation}\label{zdot}
z(t_{obs})=\frac{a(t_{obs})}{a(t_{s})}-1,
\end{equation}
 where $t_{s}$  is the time at which the source emitted the
radiation and $t_{obs}$ is the time of observation. In
 expression (\ref{zdot}), any peculiar motion of the object is ignored. After the time interval $\Delta t_{obs}$ the source redshift becomes
\begin{eqnarray}
z(t_{obs}+\Delta t_{obs})=\frac{a(t_{obs}+\Delta
t_{obs})}{a(t_{s}+\Delta t_{s})}-1 ,
\end{eqnarray}
where $\Delta t_{s}$ is time interval of the source emitted.
One can write the first order approximation of the above equation as
\begin{eqnarray}
\frac{\Delta z}{\Delta
t_{obs}}\simeq\frac{(\dot{a}(t_{obs})-\dot{a}(t_{s}))}{a(t_{s})},
\end{eqnarray}
or in term of the Hubble parameter:
\begin{equation}\label{zdot2}
 \dot{z}=H_{0}[1+z-\frac{H(z)}{H_{0}}].
\end{equation}
Equation (\ref{zdot2}) is also known as McVittie equation \cite{McVittie}. It
clearly shows that $\dot{z}$ traces H(z), the Hubble parameter
at redshift z. As stated in the introduction,  $\dot{z}$ measures the rate of
expansion of the universe: $\dot{z} > 0 $ ($< 0 $) indicates the
accelerated (decelerated) expansion of the universe, respectively.
For a coasting universe $\dot{z}= 0$. The redshift variation is
related to the apparent velocity shift of the source:
\begin{equation}\label{dv}
 \Delta v=c\frac{\Delta z}{1+z}.
\end{equation}
Thus, using equation (\ref{zdot2}) and (\ref{dv}) one can obtain,
\begin{equation}\label{drift}
\dot{v}=\frac{cH_{0}}{1+z}[1+z-\frac{H(z)}{H_{0}}],
\end{equation}
where $\dot{v}=\frac{\Delta v}{\Delta t_{obs}}$ and $H_{0} =100 h Km/sec/Mpc $. In the
standard cosmological model $(\Lambda CDM)$, the change in redshift for a time interval
of $\Delta t_{obs} = 10 yr$ is $\Delta z\simeq 10^{-9}$.
For a source at redshift $z = 3$, the corresponding velocity shift is of the order of $ \Delta v \simeq 7.5 cm/s$. To measure
this weak signal, the author in \cite{Loeb} pointed out the detection of signal of such a tiny magnitude
might be possible by observation of
the $LY\alpha$ forest in the QSO spectrum for a decade \cite{Jain}.

To observe such a tiny signals, a new generation of Extremely Large
Telescope (ELT), equipped with a
high resolution, extremely stable and ultra high precision
spectrograph is needed. Using the Cosmic Dynamics
Experiment (CODEX) operation and performing Monte Carlo simulations of quasars absorption
spectra \cite{Pasquin,Pasquini} one obtains the $\dot{z}$ measurements.

In the following, we use three sets of data (8 points) for redshift drift generated by performed Monte Carlo. We investigate various models of the
universe which can explain late time acceleration,
and  compare the chameleon scalar field model with these models and observational data \cite{Lis}, \cite{Pasquin}--\cite{Liske}.

\subsection{$\Lambda CDM $ model}

One of the widely studied model of dark energy is
XCDM parametrization in which the dark energy is
characterized by  time independent EoS parameter, with $p=\omega_{x}\rho$. Here $\Lambda CDM$ model
means standard cosmological model with $\omega=-1$, $\Omega_{m0} = 0.3$ and $\Omega_{\Lambda0} = 0.7$.  For
acceleration in this model we need $\omega_{x}<\frac{-1}{3}$. The Friedmann
equation in this model for a flat universe is given by
\begin{equation}
 \Big[\frac{H(z)}{H_{0}}\Big]^{2}=\Omega_{m}(1+z)^{3}+\Omega_{x}(1+z)^{3(1+\omega_{x})},
 \end{equation}
where $\Omega_{m}$ and $\Omega_{x}$
are the fractional matter and dark energy
densities at the present epoch respectively with $\Omega_{m}+\Omega_{x}=1$.

\subsection{CPL model}

Another popular parametrization which explains evolution
of dark energy is the CPL model \cite{Chevalier} in which in a flat universe the time varying EoS parameter is parametrized by,
\begin{equation}
\omega(z)=\omega_{0}+\omega_{1}(\frac{z}{1+z}).
\end{equation}
The Hubble parameter in the model is given by,
\begin{equation}
 \Big[\frac{H(z)}{H_{0}}\Big]^{2}=\Omega_{m}(1+z)^{3}+(1-\Omega_{m})(1+z)^{3(1+\omega_{0}
 +\omega_{1})}\times\exp\Big[-3\omega_{1}(\frac{z}{1+z})\Big].
 \end{equation}
The parametrization is fitted for different values of $\omega_0$ and $\omega_1$. The velocity drift with respect to the source
redshift is shown in figure (4) \cite{Lis1}. As can be seen, the best fit values are for $\omega_0=-2.4$ and $\omega_1=3.4$.\\

\begin{tabular*}{2. cm}{cc}
\includegraphics[scale=.75]{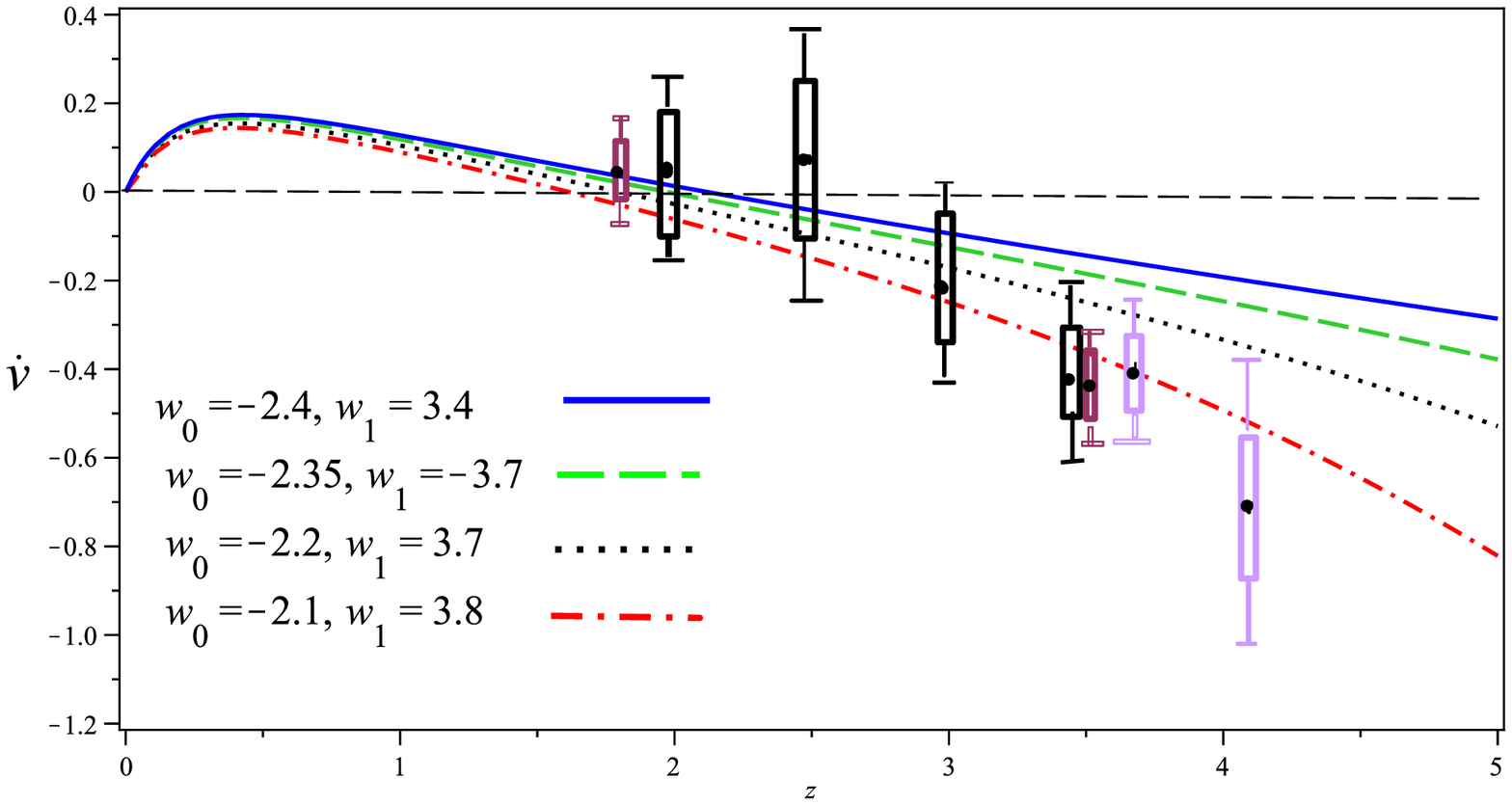}\\
Fig.4: The plot of the velocity drift ($h_{100}cm s^{-1} yr^{-1}$) versus $z$ for CPL parametrization \\ with $\Omega_{m}=0.3$
\end{tabular*}\\

\subsection{Chameleon model}

By using $a=\frac{a_{0}}{1+z}$, and taking derivative we get
\begin{eqnarray}\label{hdot2}
\frac{dH(t)}{dt}=-(1+z)H(z)\frac{dH(z)}{dz}.
\end{eqnarray}
From equations (\ref{fried1}), (\ref{fried2}) and (\ref{hdot2}) we obtain the following expression
for EoS parameter,
\begin{eqnarray}\label{omegaeff}
\omega_{eff}=-1+\frac{(1+z)r^{(1)}}{3r},
\end{eqnarray}
where $r= \frac{H^{2}}{H^{2}_{0}}$ and
$r^{(n)}=\frac{d^{n}r}{dz^{n}}$. In figure (5), it can be seen that for different values of $V_0$, the EoS parameter crosses $-1$ at different $z$. It shows that for larger values of $V_0$, the phantom crossing is more appropriate with the observational data \cite{Sur}.\\

\begin{tabular*}{2. cm}{cc}
\includegraphics[scale=.75]{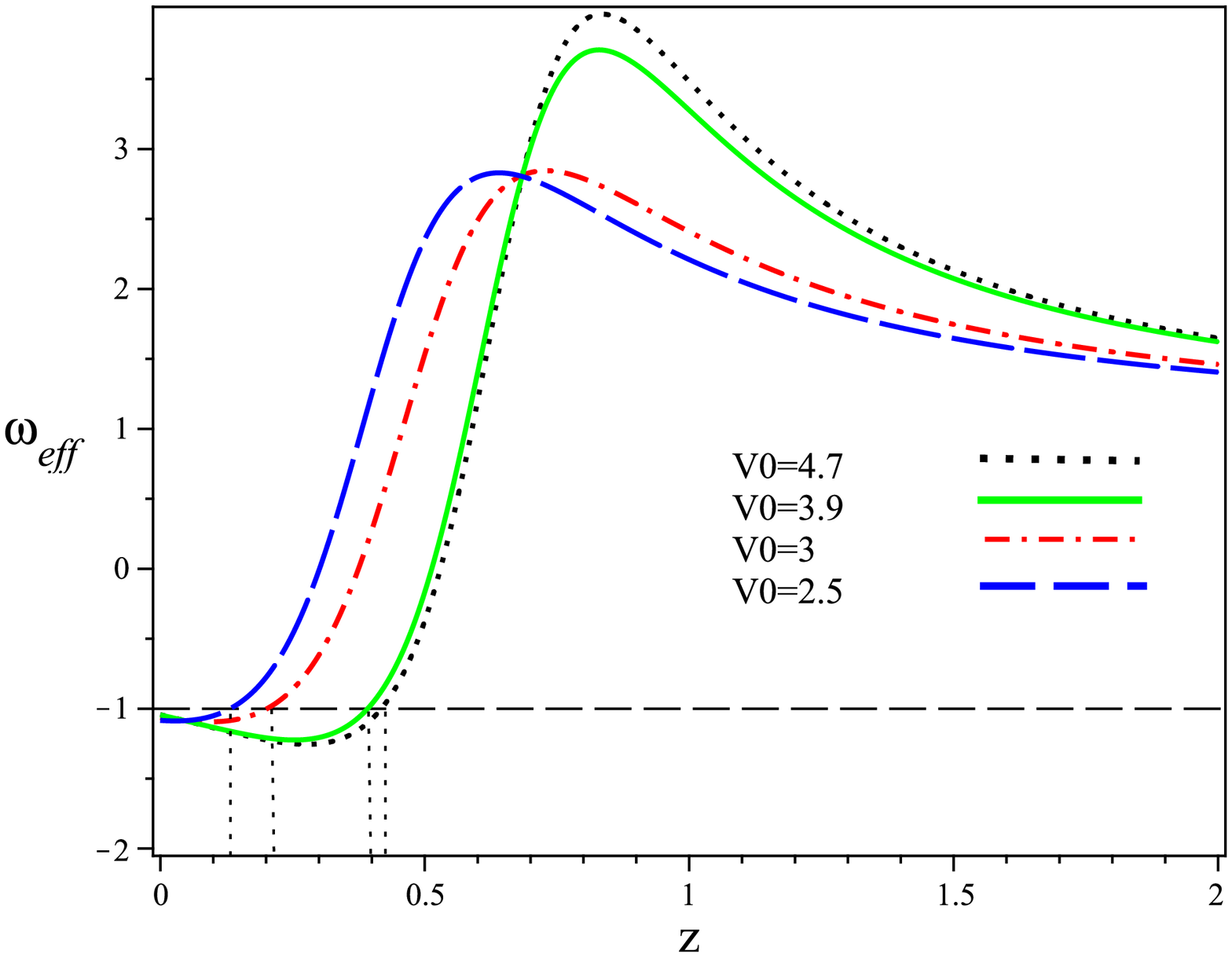}\\
Fig.5: Plot of $\omega_{eff}(z)$
with $f={\it f_{0}}\,{{\rm e}^{b\phi\left(z \right)
 }}$and $V=V_{0}\phi(z)^{n}$
 for fixed value $f_{0}=-7$.\\($b = -1$ ,$n=-1$). Initial values are $\phi(0)=1$, $\dot{\phi}(0)=-0.8$.\\
\end{tabular*}\\

It also shows that $\omega_{eff}$ will decrease before crossing along with the increase of redshift $z$. This result implies
that the behavior of dark energy is different in different slices of redshifts, and inspires us to
separate redshifts into several pieces and to investigate each piecewise separately which can be investigated in another work.

From equation (\ref{omegaeff}) we also obtain,
\begin{eqnarray}
\Big[\frac{H(z)}{H_{0}}\Big]^{2}=\exp\Big[{3\int^{z}_{0}\frac{1+\omega_{eff}(\tilde{z})}{1+\tilde{z}}d\tilde{z}}\Big]\cdot
\end{eqnarray}
 From numerical computation of $\omega_{eff}(z)$, one can obtain $H(z)$. Then, using equation (\ref{drift}), one leads to the velocity drift. Figure (6) shows the velocity drift against $z$ for various potential functions. In comparison with the CPL model our model is in a better agreement with the experimental data for $z>3$ whereas the CPL model is more appropriate for $z<3$.\\

\begin{tabular*}{2. cm}{cc}
\includegraphics[scale=.75]{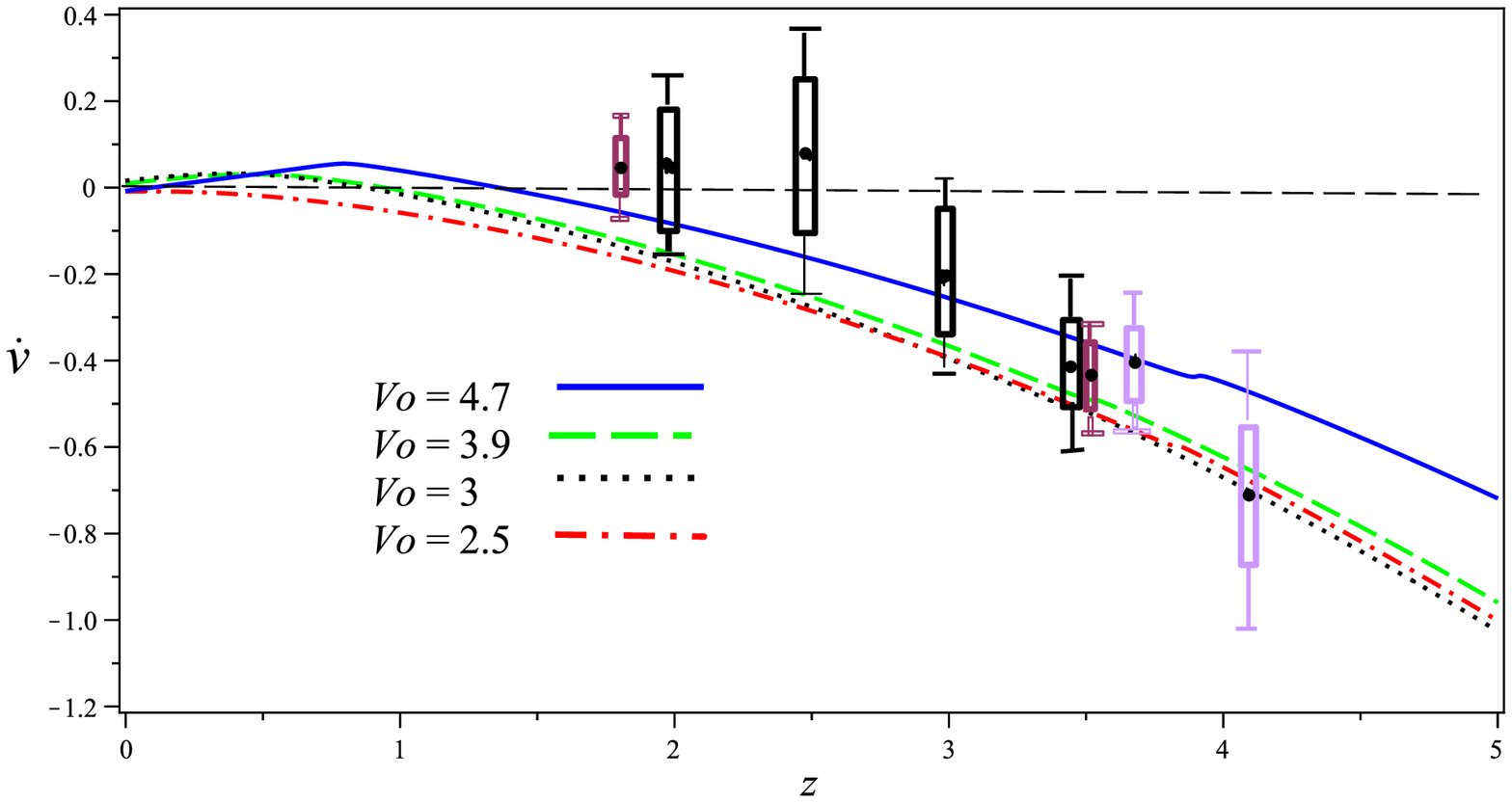}\hspace{.5 cm}\\
Fig. 6 : Plot of   $\dot{v}(z)$
with $f={\it f_{0}}\,{{\rm e}^{b\phi\left(z \right)
 }}$and $V=V_{0}\phi(z)^{n}$
 for $f_{0}=-7$,($b = -1$ ,$n=-1$).\\Initial values are $\phi(0)=1$,
$\dot{\phi}(0)=-0.8$.
\end{tabular*}\\

\section{Conclusion}

In this paper, we study the evolution of the gravitational and scalar fields in chameleon cosmological model in which a light scalar field (chameleon field) nonminimally coupled to the matter Lagrangian. We find that the evolution of the scale factor of the universe is non-singular in a bouncing cosmology, with an initial contracting phase which lasts until to a non-vanishing
minimal radius is reached and then smoothly transits into an expanding phase which provides a possible solution
to the singularity problem of standard Big Bang cosmology. In addition, The evolution of the cosmological EoS parameter, with a transition from $\omega >-1$ in the past to the $\omega<-1$ in the recent past is favored for negative $f(\phi)$, in agreement with the current observational data.
 The phantom crossing occurs for different values of $V_0$ within the range of observationally accepted redshift $z$ ($0.1<z<0.45$).

We then analyze the chameleon model with the CRD test. The variation of
velocity drift for different values of $V_0$ with redshift
$z$ is shown in figure (6). A comparison between this model with CPL model shows that it is in better agreement with the experimental data for values of redshift $z$ greater than $3$. We also obtain that the Hubble
parameter, H(z), in the model is potential dependent across the redshift. In order to better constrain the parameters of the model we need more redshift drift data in the redshift range $z<1.9$.

{\bf Acknowledgement}

The authors would like to thank the anonymous reviewers for their careful review and helpful comments. This work was partially supported by the University of Guilan Grant Committee.

\end{document}